\documentclass[%
 reprint,
 amsmath,amssymb,
 aps,
]{revtex4-1}

\usepackage{graphicx}
\usepackage{dcolumn}
\usepackage{bm}

\usepackage{subfigure}
\usepackage{xcolor}

\begin{document}


\title{Thermodynamic Behaviour of Magnetized QGP within the Self-Consistent Quasiparticle Model}

\author{Sebastian Koothottil}
\email{sebastian$_$dop@uoc.ac.in}
 \altaffiliation[Also at ]{Department of Physics, University of Calicut.}
\author{Vishnu M. Bannur}%
\affiliation{%
Department of Physics, University of Calicut, Kerala-673635, India 
}%





\begin{abstract}
The self-consistent quasiparticle model has been successful in studying QCD thermodynamics. In this model, the medium effects are taken into account by considering quarks and gluons as quasiparticles with temperature dependent masses which are proportional to the plasma frequency. The present work involves the extension of this model in the presence of magnetic fields. We have included the effect of the magnetic field by considering relativistic Landau Levels. The quasiparticle masses are then found to be dependent on both temperature and magnetic field. The thermo-magnetic mass thus defined allows obtaining the thermodynamics of magnetised quark matter within the self-consistent quasiparticle model. The model then has been applied to the case of 2-flavor Quark-Gluon Plasma and the equation of state obtained in the presence of magnetic fields. 
\end{abstract}

\pacs{Valid PACS appear here}
\maketitle


\section{Introduction}
High energy collisions have succeeded in recreating the state of matter called Quark Gluon Plasma (QGP) which is believed to have existed shortly after the big bang. It has been observed that QGP produced in high energy collisions behave very much like a nearly perfect fluid \cite{back1,*adams1,*adcox1,*arsene1}. These collisions mostly occur with a finite impact parameter. During off central collisions, the charged ions thus can produce very large magnetic fields reaching up to $eB\approx (1-5) m_{\pi}^2$ \cite{skokov3,zhong4}. These magnetic fields may exist only for a short time but, depending on the transport coefficients, they may reach their maximum value and can be stationary during this time \cite{tuchinstatic1,marasinghetuchinstatic,tuchin,fukushimastatic}.  The magnetic field can cause different phenomena like magnetic catalysis \cite{catalysis1,catalysis2,Gusynin1996, Fukushima2013}, chiral magnetic effect \cite{chiral1,chiral2,chiral3}, etc. in the QGP. The equation of state is important for studying the particle spectra created in heavy-ion collisions.  Very strong magnetic fields are estimated to have existed right after the big bang \cite{Grasso2001}. Effects of external magnetic fields are relevant in the context of strongly magnetised neutron stars too \cite{Duncan1992,*Broderick2000,*Cardall2001,*Rabhi2008,*Rabhi2011}. Therefore, it is of importance to investigate the behaviour of QGP under magnetic fields, in particular, the effect on the QCD thermodynamics \cite{Strickland2012, Avancini2018}. There have been several investigations as to how these magnetic fields affect the transport coefficients \cite{Kurian2018, Kurian2018a}. 
  
    In this work, we intend to understand the effect of magnetic fields on QCD thermodynamics and obtain the equation of state by extending a quasiparticle model for QGP, called the self-consistent quasiparticle model. We incorporate the effect of the magnetic field by modifying the thermal mass using the relativistic Landau Levels.  Such modified masses in the presence of magnetic fields can be used to calculate thermodynamic quantities like energy density, pressure, entropy density,  etc. Besides, by applying this formalism to the 2-flavor system, we obtain its thermodynamics in the presence of background magnetic field and examine the qualitative behaviour of the equation of state.   
\section{The self-consistent quasiparticle model}
In quasiparticle models, the thermal properties of interacting real particles are modelled by noninteracting quasiparticles.  The quasiparticles have an effective mass which is determined by the collective properties of the medium \cite{Goloviznin1993, Peshier1994}.  Such models include the Nambu-Jona-Lasinio and PNJL based quasiparticle models \cite{Dumitru2002, *Fukushima2004,*Ghosh2006,*Abuki2009,*Tsai2009} or those that include effective mass with the Polyakov loop \cite{Elia1997,*Castorina2007,*Castorina2007a}. There are also quasiparticle models based on Gribov-Zwanziger quantisation \cite{su2015}. Other effective mass quasiparticle models include self-consistent and single parameter quasiparticle models \cite{Bannur2007,Bannur2007a,Bannur2007b,Bannur2007c,Bannur2008}.  There are quasiparticle models which incorporate the medium effects by considering quasiparticles with effective fugacities too. Such models have been quite successful in describing the lattice QCD results \cite{Chandra2009, Chandra2011}.

     The self-consistent quasiparticle model here considers QGP as consisting of non-interacting quasiparticles with effective masses which depend on thermodynamic quantities and encode all medium interactions \cite{Bannur2007, Bannur2008}.  Since the thermal mass depends on thermodynamic quantities which in turn depends on thermal mass, the whole problem is solved self-consistently.

In the self-consistent model, following standard thermodynamics, all thermodynamic quantities are derived from the expressions for energy density and number density. The expression for energy density is, 
\begin{equation}
\varepsilon=\frac{g_f}{2 \pi^2} \int_0^\infty dk\; k^2 \frac{\omega_k}{z^{-1} e^{\frac{\omega_k}{T}} \mp 1}, \label{eq:energydensity}
\end{equation}
where $g_f$ is the degeneracy and $\mp$ refers to bosons and fermions. $z$ is the fugacity.  The expression for number density is, 
\begin{equation}
n_{g/q}=\frac{g_f}{2 \pi^2}\int_0^\infty dk\; k^2 \frac{1}{z^{-1} e^{\frac{\omega_k}{T}} \mp 1}. \label{eq:numberdensity}
\end{equation}

The single particle energy $\omega_k$ is approximated to a simple form, 
\begin{equation}
\omega_k = \sqrt{k^2 + m_g^2}
\end{equation}
and 
\begin{equation}
\omega_k = \sqrt{k^2+m_q^2},
\end{equation}
for gluons and quarks respectively.  This approximation is valid at high temperatures only.

It is known that quasiparticles acquire "thermal mass" of order $gT$ at one loop order \cite{Pisarski1989, Pisarski1993, Peshier1996}.    In the model that we study here, the thermal mass is defined as proportional to the plasma frequencies as, 
\begin{equation}
m_g^2 = \frac{3}{2} \omega_p^2 \;\;\;\text{and}\;\;\; m_q^2 = 2 m_f^2,\label{eq:massplasmafreq}
\end{equation}
for massless particles. For massive quarks $m_q^2$ is written as, 
\begin{equation}
m_q^2 = (m_0 + m_f)^2 + m_f^2.\label{eq:mqmassive}
\end{equation}
 The plasma frequencies are calculated from the density dependent expressions
 \begin{equation}
 \omega_p^2 = a_g^2 g^2\frac{ n_g}{T} + a_q^2 g^2 \frac{n_q}{T},\label{eq:plasmafrequencygluons}
 \end{equation}
 for gluons and,
 \begin{equation}
 m_f^2 = b_q^2 g^2 \frac{n_q}{T} ,\label{eq:plasmafrequencyquarks}
 \end{equation}
 for quarks. Here $n_q$ is the quark number density and $n_g$ is is the gluon number density. $g^2=4 \pi \alpha_s$ is the QCD running coupling constant. The coefficients $a_g$, $a_q$, $bq$ are determined by demanding that as $T\rightarrow \infty$, $\omega_p$ and $m_f$ both go to the corresponding perturbative results.   The motivation for choosing such an expression for plasma  frequency is that the plasma frequency for electron-positron plasma is known to be proportional to $n/T$ in the relativistic limit \cite{Medvedev1999, Bannur2006}. Since the thermal masses appear in the expression for the density, we need to solve the density equation self-consistently to obtain the thermal mass, which may be used to evaluate the thermodynamic quantities of interest.  The result obtained have shown a good fit with lattice data even at temperatures near $T_c$ \cite{Bannur2012}.
\section{Extension of the self-consistent model in magnetic field}
In this section, we extend the self-consistent quasiparticle model to a system with zero chemical potential, in the presence of magnetic fields. The quantisation of fermionic theory in magnetic fields has been known.  The energy eigenvalues are obtained as Landau levels and have been subject to several investigations \cite{Bhattacharya2007, Bruckmann2018}.  In the Landau gauge $A_y = Bx$ so that $\textbf{B} = B \hat{z}$ and the energy eigenvalues are obtained as, 
  \begin{equation}
  E_j=\sqrt{m^2+k_z^2+2 j \mid q_f eB\mid}. \label{eq:energyeigenvalueslandaulevel}
  \end{equation}
  Here, $q_f e$ is the charge of the fermion and $j=0,1,2,..$ are the Landau energy levels.

   In the presence of a magnetic field $B$, the integral over the phase space is modified as \cite{Fraga2012,Mizher2010, Chakrabarty1996, Bruckmann2017, Tawfik2016}, 
 
  \begin{equation}
  \int \frac{d^3k}{(2 \pi)^3}\rightarrow \frac{\mid q_f eB \mid}{2 \pi}\sum_{j=0}^{\infty} \int \frac{dk_z}{2 \pi}\left(2-\delta_{0j}
  \right),\label{eq:dimensionalreduction}
  \end{equation} 
  where, $(2-\delta_{0j})$ is the degeneracy of the $j^{th}$  Landau Level \cite{Kohri2004}.
 \subsection{Thermo-Magnetic Mass for Quarks}
 Making use of equations \eqref{eq:energyeigenvalueslandaulevel} and \eqref{eq:dimensionalreduction}, we can write the expression for number density in the presence of magnetic fields, for a system at zero chemical potential as, 
 \begin{align}
 n_q &{}= \frac{g_f q_f eB}{(2 \pi)^2}\sum_{j=0}^{\infty} \Bigg[ 2  \int_{-\infty}^\infty dk_z \frac{1}{e^{\sqrt{(\frac{k_z}{T})^2 + (\frac{m_{qj}}{T})^2}} +1}\nonumber \\& -\int_{-\infty}^\infty dk_z \frac{1}{e^{\sqrt{(\frac{k_z}{T})^2 + (\frac{m_{qj}}{T})^2}} +1} \delta_{0j}\Bigg],\label{eq:numberdensitymagfield}
 \end{align} 
 where, we have assigned for simplicity, 
 \begin{equation}
 m_{q_j}^2 = m_q^2 + 2 j |q_f e B|.\label{eq:mqj}
 \end{equation}
 Equation \eqref{eq:numberdensitymagfield} can be further simplified to 
 \begin{align}
T^2 F_q^2 =&{} \sum_{j=0}^{\infty} \Bigg(2  \sum_{l=1}^{\infty} (-1)^{(l-1)} \frac{m_{qj}}{T} K_1\left(l \frac{m_{qj}}{T}\right) \nonumber \\& - \sum_{l=1}^{\infty} (-1)^{(l-1)} \frac{m_{qj}}{T} K_1\left(l \frac{m_{qj}}{T} \right)\delta_{0j}\Bigg).\label{eq:1}
 \end{align}
 Here we have defined for later convenience, 
 \begin{equation}
 F_q^2 = \frac{(2 \pi)^2}{2 g_f |q_f eB|}\frac{n_q}{T^3}.
 \end{equation}

 For massive quarks, $m_q$ can be obtained from equation \eqref{eq:mqmassive}, with $m_f$ calculated just as in \cite{Bannur2008},  
 
 \begin{equation}
 m_f^2 = c_q^2 g^2 \frac{n_q}{T}.
 \end{equation}
 
 Or, 
 \begin{align}
 \left(\frac{m_f}{T} \right)^2 &= c_q^2 g^2 \frac{n_q}{T^3} \nonumber\\
                        &= \bar{c}_q ^2 F_q^2, \label{eq:mfbyT}
   \end{align}
   where, 
   \begin{equation}
   \bar{c}_q^2 =  2 c_q^2 g^2\frac{g_f |q_f e B|}{(2 \pi)^2}. 
   \end{equation}
   Combining equations \eqref{eq:mqmassive}, \eqref{eq:mqj}, and \eqref{eq:mfbyT}, we can write, 
   \begin{equation}
   \left(\frac{m_{qj}}{T}\right)^2 = \left[\frac{m_0}{T}+ \bar{c}_q Fq\right]^2 + \bar{c}_q^2 F_q^2  + 2j \frac{|q_f eB|}{T^2}. \label{eq:mqjinFq}
   \end{equation}
   Using \eqref{eq:mqjinFq} in equation \eqref{eq:1}, and simplifying the Kronecker delta we get ,
    \begin{widetext}
    \begin{align} 
  T^2 F_q^2&=\sum_{l=1}^{\infty} (-1)^{(l-1)} \Bigg[2\sum_{j=0}^{\infty}  \sqrt{\left[\frac{m_0}{T}+ \bar{c}_q Fq\right]^2 + \bar{c}_q^2 F_q^2  + 2j \frac{|q_f eB|}{T^2} }  K_1\left(l \sqrt{\left[\frac{m_0}{T}+ \bar{c}_q Fq\right]^2 + \bar{c}_q^2 F_q^2  + 2j \frac{|q_f eB|}{T^2} }\right)\nonumber\\ &- \sqrt{\left[\frac{m_0}{T}+ \bar{c}_q Fq\right]^2 + \bar{c}_q^2 F_q^2  } \;  K_1\left(l \sqrt{\left[\frac{m_0}{T}+ \bar{c}_q Fq\right]^2 + \bar{c}_q^2 F_q^2  }\right)\Bigg], \label{eq:Fq}
  \end{align}
\end{widetext}
    
   where $K_n(x)$ are the modified Bessel functions of the second kind.  Solving this equation for $F_q$ and using equation \eqref{eq:mqjinFq} we get the thermo-magnetic mass which depends both on temperature and magnetic field. This thermo-magnetic mass can be used to obtain the thermodynamics of the system considering it as a collection of non-interacting quasiparticles with mass depending on temperature and magnetic field.
    \subsection{Thermo-Magnetic Mass for Gluons}
The density-dependent expression for plasma frequency for gluons is given by equation \eqref{eq:plasmafrequencygluons}. The expression for gluon number density $n_g$ remains unchanged because gluons are chargeless and are not directly affected by magnetic fields. They are only indirectly affected by their coupling to the electrically charged quarks. Thus the term $n_q$ changes as explained and so the gluons also acquire a thermo-magnetic mass.  We have, 
   \begin{align}
   n_g = \frac{g_g}{2 \pi^2} \int_0^\infty dk \; k^2 \frac{1}{e^{\omega_k/T}-1}. \label{eq:gluonnumberdensityeq}
   \end{align}
 Following \eqref{eq:plasmafrequencygluons}, we define the plasma frequency as, 
 \begin{equation}
  \omega_p^2 = a_g^2 g^2\frac{ n_g}{T} + d_q^2 g^2 \frac{n_q}{T}. 
 \end{equation} 
 Making use of \eqref{eq:massplasmafreq} and simplifying equation \eqref{eq:gluonnumberdensityeq},
  \begin{align} 
   f_g^2&=(\bar{a}_g^2 f_g^2+\bar{d}_q^2 F_q^2)\sum_{l=1}^\infty \frac{1}{l}K_2[l(\bar{a}_g^2 f_g^2+\bar{d}_q^2 F_q^2)^{1/2}]. \label{eq:solveforfg}
   \end{align}
    where, 
    \begin{equation}
   f_g^2 = \frac{2 \pi^2}{g_g} \frac{n_g}{T^3},
   \end{equation}
   \begin{align}
    \bar{a}_g^2 &= \frac{3}{4 \pi^2} g_g a_g^2 g^2, \\
   \bar{d}_q^2 &= \frac{3}{2} \frac{a_q^2 g^2}{(2 \pi)^2} 2 g_f |q_f eB|,
   \end{align}
   and $F_q $ is obtained as solution to \eqref{eq:Fq}.
  Now, solving equation \eqref{eq:solveforfg} for $f_g$ using the above, we obtain the thermo-magnetic mass using, 
  \begin{equation}
  \left(\frac{m_g}{T} \right)^2 = \bar{a}_g^2 f_g^2 + \bar{d}_q^2 F_q^2.
  \end{equation}

   The coefficients $c_q$, $a_g$ and $d_q$ are determined by demanding that as $T\rightarrow \infty$ the expression for frequency approaches the corresponding perturbative QCD results. 
   
   \subsection{QCD Thermodynamics in Background Magnetic Field }
   \subsubsection{Energy Density}
We start with the energy density. Using equations \eqref{eq:energydensity},  \eqref{eq:energyeigenvalueslandaulevel} and \eqref{eq:dimensionalreduction} the contribution to the energy density from quarks becomes,
  \begin{align}
  \varepsilon_q &= \frac{12 n_q |q_f e B|}{4 \pi^2} \sum_{j=0}^{\infty} \Bigg[ 2 \int_{-\infty} ^{\infty} dk_z \frac{\omega_{k_{zj}}}{e^{\omega_{k_{zj}/T}}+1} \nonumber\\ &\;-\int_{-\infty} ^\infty dk_z \frac{\omega_{k_{zj}}}{e^{\omega_{k_{zj}}/T}+1} \delta_{0j}\Bigg],\label{eq:energydensitymagfield1}
  \end{align}
  where, 
  \begin{align}
  \omega_{k_{zj}} &= \sqrt{m_q^2 + k_z^2+2 j |q_f e B|} \nonumber\\
                  &= \sqrt{m^2_{q_j}+ k_z^2}.
  \end{align}

  Equation \eqref{eq:energydensitymagfield1} becomes, after some algebra,
 
  \begin{align}
  \varepsilon_q = {}& \frac{12 n_q |q_f eB|}{2 \pi^2}\frac{T^2}{2}\sum_{l=0}^\infty \frac{(-1)^{(l-1)}}{l^2}\Bigg\{2\sum_{j=0}^\infty \nonumber\\&\Bigg[ \left(\beta m_{q_j}l \right)K_1\left(\beta m_{q_j}l\right) + \left(\beta m_{q_j} l \right)^2K_0\left(\beta m_{q_j}l\right)\Bigg]\nonumber\\
&-\left[ \left(\beta {m_{q}l}\right)K_1\left( \beta {m_{q}l}\right)+ \left(\beta {m_{q}l}\right)^2K_0\left(\beta {m_{q}l}\right)\right]\Bigg\}.
\end{align}

The expression for the energy density of gluons remains unchanged as they are chargeless. The energy density is indirectly affected by the interaction with quarks, included through the thermo-magnetic mass of gluons.
\begin{align}
\varepsilon_g =&{}
\frac{g_g T^4}{2\pi^2} \sum_{l=1}^\infty \frac{1}{l^4}\Bigg[\left(\beta m_g l\right)^3 K_1\left(\beta m_g l\right)\nonumber \\ &+ 3 \left(\beta m_g l\right)^2 K_2\left(\beta m_g l\right)\Bigg].
\end{align}
The contribution to energy density from quarks and gluons is obtained as 
\begin{equation}
\varepsilon = \varepsilon_g + \sum_q \varepsilon_q.
\end{equation} 
\subsubsection{Pressure}

It has been known that the presence of magnetic fields breaks the $O(3)$ rotational symmetry resulting in a pressure anisotropy \cite{Ferrer2010,  Kohri2004, Martinez2003, Chaichian2000}. There have also been arguments suggesting that the total pressure is indeed isotropic and the issue has been subject to some debate \cite{Potekhin2012,  Ferrer2012, Dexheimer2013}. The scheme dependence of pressure anisotropy has been discussed in \cite{Bali2014, Bali2013}. Here the authors have distinguished between two schemes. The $B-$scheme, which corresponds to a setup in which the magnetic field $B$ is kept constant during the compression, and the $\Phi$-scheme corresponding to a setup in which the magnetic flux is kept constant during the compression. They showed that pressure anisotropy appears only in the $\Phi$-scheme, i.e., $P_{\perp}^{(B)}=P^{(B)}_{\parallel}$ and $P_{\perp}^{(\Phi)}\neq P^{(\Phi)}$, where $P_{\perp}$and $P$ denote the transverse and longitudinal pressure respectively.  They also showed that the longitudinal pressure is scheme independent. Thus, $P^{(\Phi)}_{\parallel}  = P^{(B)}_{\parallel}$. In the $\Phi$ scheme the longitudinal and transverse pressures were found to be related by,
   \begin{equation}
   P_{\perp}^{\Phi} = P_{\parallel} - e M\cdot B.\label{eq:transversepressure}
   \end{equation} 
  The magnetization can be calculated in the canonical ensemble using the equation,
   \begin{equation}
   M = \frac{T}{V} \frac{1}{e} \frac{\partial \ln Z}{\partial B},\label{eq:magnetization}
   \end{equation}
  where $Z$ is the partition function.

The contribution to $P_{\parallel}$ from quarks and gluons can be calculated from the expression for thermodynamic pressure in the self-consistent quasiparticle model \cite{Bannur2007a}. We will denote this as $P$. 
\begin{align}
\frac{P_{g/q}}{T} ={}& \mp \frac{g_f}{8 \pi^3} \int_0^\infty d^3k \ln\left(1 \mp e^{-\beta \epsilon_k}\right) \nonumber \\ {}&+ \int d\beta\; \beta \frac{g_f}{2 \pi^2} m \frac{dm}{d\beta}\;\int_0 ^\infty d^3k \frac{k^2}{\epsilon_k \left(e^{\beta \epsilon_k} \mp 1\right)},\label{eq:Pressurequarkgluon}
\end{align}
where $\mp$ for bosons and fermions respectively. 
 The contribution to the thermodynamic pressure from quarks can be obtained by changing the integral and energy eigenvalues according to equations \eqref{eq:dimensionalreduction} and \eqref{eq:energyeigenvalueslandaulevel}. Making these changes and simplifying the integral we obtain,
\begin{align}
\frac{P_{q} }{T} =&{} \frac{g_f |q_f e B|}{2 \pi^2}\sum_{l=1}^{\infty} (-1)^{l-1} \Bigg\{ 2 \sum_{j=0}^{\infty}\Bigg[\frac{1}{\beta l^2} (\beta m_{q_j} l) K_1(\beta m_{q_j} l) \nonumber\\{}{}{}{}{}{}{}{}{}&+ \int d\beta\; \beta m_{q_j} \frac{\partial m_{q_j}}{\partial \beta} K_0(\beta m_{q_j} l) \Bigg]\nonumber \\
-&{}\Bigg[\frac{1}{\beta l^2}(\beta m_q l) K_1(\beta m_q l)+\int d\beta \; \beta m_q \frac{\partial m_q}{\partial \beta} K_0(\beta m_q l)\Bigg]\Bigg\}. \label{eq:quarkpressuremagfield}
\end{align}

The contribution to the pressure from gluons can be obtained by replacing the thermal mass by thermo-magnetic mass.  From equation \eqref{eq:Pressurequarkgluon}, we can write the contribution from gluons as, 
\begin{align}
\frac{P_{g}}{T} =&{} \frac{g_f}{2 \pi^2}\sum_{l=1}^\infty \frac{1}{l^4} \Big[  T^3 (\beta m_g l)^2 K_2(\beta m_g l)\nonumber\\ &{}+\int d\beta\; \frac{\beta^3}{m_g} \frac{\partial m_g}{\partial \beta}  (\beta m_g l)^3 K_1(\beta m_g l) \Big].\label{eq:Pressuregluonsmagfield}
\end{align}

It can easily be shown that the expressions for thermodynamic pressure in\eqref{eq:Pressuregluonsmagfield} and \eqref{eq:quarkpressuremagfield} are related to the corresponding expressions for energy densities as
\begin{equation}
\varepsilon = T \frac{\partial P}{\partial T} - P,\label{eq:thermodynamicconsistency}
\end{equation}
ensuring thermodynamic consistency. Thus, the thermodynamic pressure can also be obtained from the energy density. Solving equation \eqref{eq:thermodynamicconsistency}, 
\begin{equation}
\frac{P}{T} = \frac{P_0}{T_0}+ \int_{T_0}^{T} dT \frac{\varepsilon}{T^2}. 
\end{equation}
Here, $P_0$ and $T_0$ are pressure and temperature at some reference points \cite{Bannur2007b}.

The entropy density can be calculated from, 
\begin{equation}
s= \frac{\varepsilon+P}{T}.
\end{equation}

   \subsubsection{The Pure-field Contribution}
   
  In addition to the contribution from magnetized matter, there are pure-field contributions to the energy density and pressure that must be taken into account\cite{Dexheimer2013, Ferrer2010, Blandford1982, Menezes2009}.  These contributions are again different in the parallel and perpendicular directions.  
   \begin{align}
   \varepsilon^{total} ={}&\varepsilon+\frac{B^2}{2}. \\
   P_{\perp}^{total}={}&P_{\perp}+\frac{B^2}{2}. \\
   P_{\parallel}^{total}={}&P - \frac{B^2}{2}.
   \end{align}

\subsection{Thermodynamics for 2-flavor System in the Presence of Magnetic Fields}
Using the above formalism, we obtain the thermodynamics for the 2-flavor system at zero chemical potential in the presence of different magnetic fields. To this end, we need a running coupling constant that depends both on temperature and magnetic field.  
 The running coupling constant needs modification in the presence of magnetic fields. It has been well known that the coupling constant is affected by magnetic fields \cite{Ferrer2015, Ayala2018}. Different ansatz for the dependence of coupling constant on magnetic fields, in the presence of magnetic fields, have been proposed \cite{ Miransky2002, Ferreira2014, Farias2014, Farias2017}. In \cite{Farias2017} the coupling constant depending on both temperature and magnetic fields was introduced in the SU(2) NJL models as, 
 \begin{equation}
 G(B,T) = c(B) \left[1-\frac{1}{1 + e^{\beta(B) [T_a(B)-T]}}\right] + s(B), \label{eq:coupling}
 \end{equation}
 where, the four parameters $c$, $\beta$, $T_a$, and $s$ were obtained by fitting the lattice data. It was shown that the thermodynamic quantities showed correct qualitative behaviour, in the presence of magnetic fields, with this parametrisation.  
 
    We make use of this ansatz for the coupling constant with the parameters as obtained in \cite{Farias2017}, to calculate the thermo-magnetic mass according to \eqref{eq:plasmafrequencygluons} and \eqref{eq:plasmafrequencyquarks} and hence obtain the thermodynamics. 
    
    Calculation of the contribution of quarks and gluons to $P_{\parallel}$ requires the value of thermodynamic pressure at some fixed temperature $T_0$. If lattice data is available $P_0$ can be chosen as the value of pressure at transition temperature $T_c$. Here we have chosen the value of thermodynamic pressure at $T_c$ from \cite{Farias2017}.
     
     For all calculations, we have taken the physical masses of quarks as in \cite{Bannur2012}.

    \begin{figure}
    \begin{center}
    \includegraphics[scale=0.6]{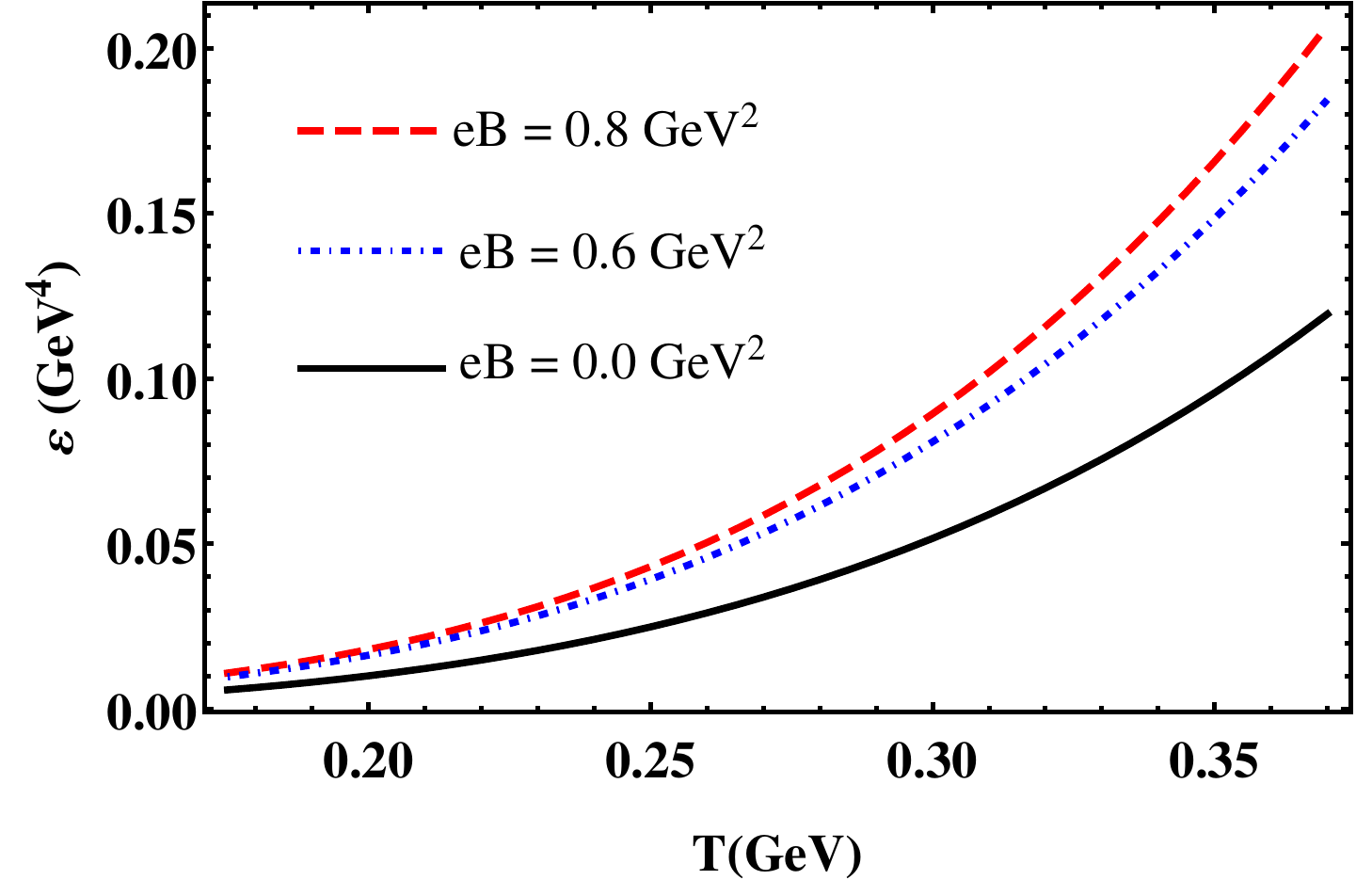}
\caption{Energy density for different magnetic fields as a function of temperature.}
\label{fig:energydensity}
\end{center}
\end{figure}
\begin{figure}
\begin{center}

\includegraphics[scale=0.6]{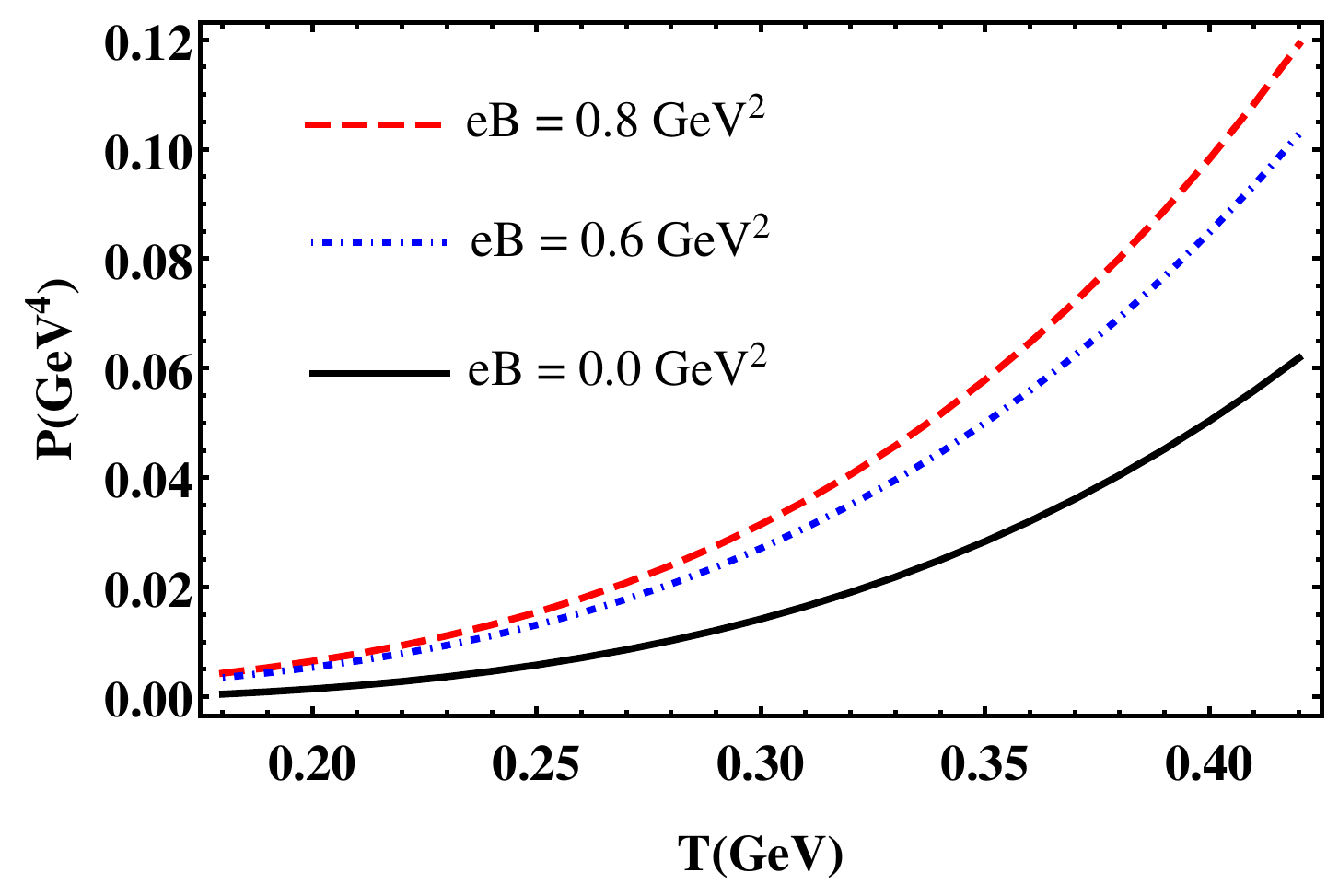}
\caption{Pressure for different magnetic fields as a function of temperature.}
\label{fig:pressure}
\end{center}
\end{figure}
\begin{figure}
\begin{center}
\includegraphics[scale=0.6]{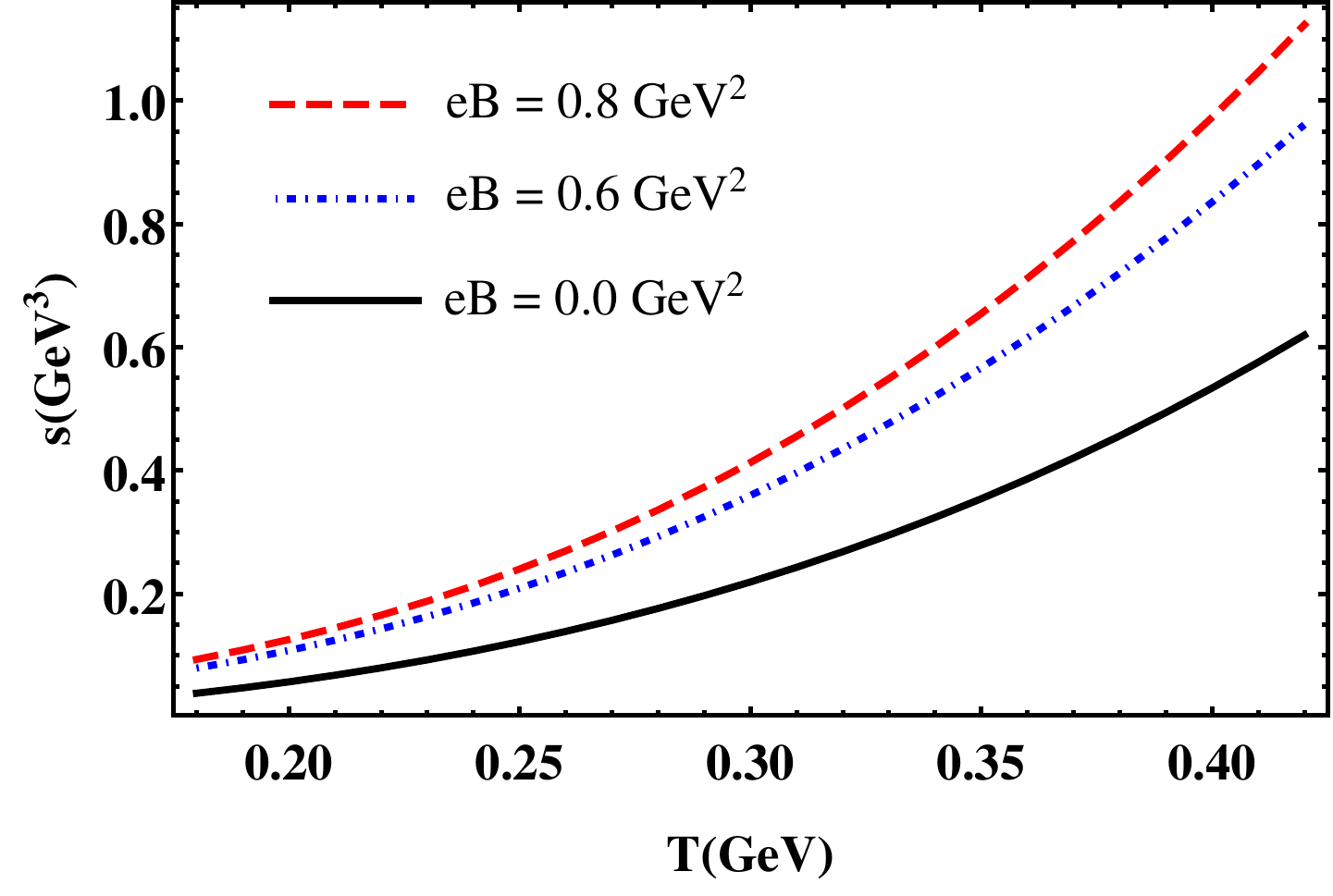}
\caption{Entropy density for different magnetic fields as a function of temperature.}
\label{fig:entropydensity}
\end{center}
\end{figure}
\begin{figure}
\begin{center}
\includegraphics[scale=0.6]{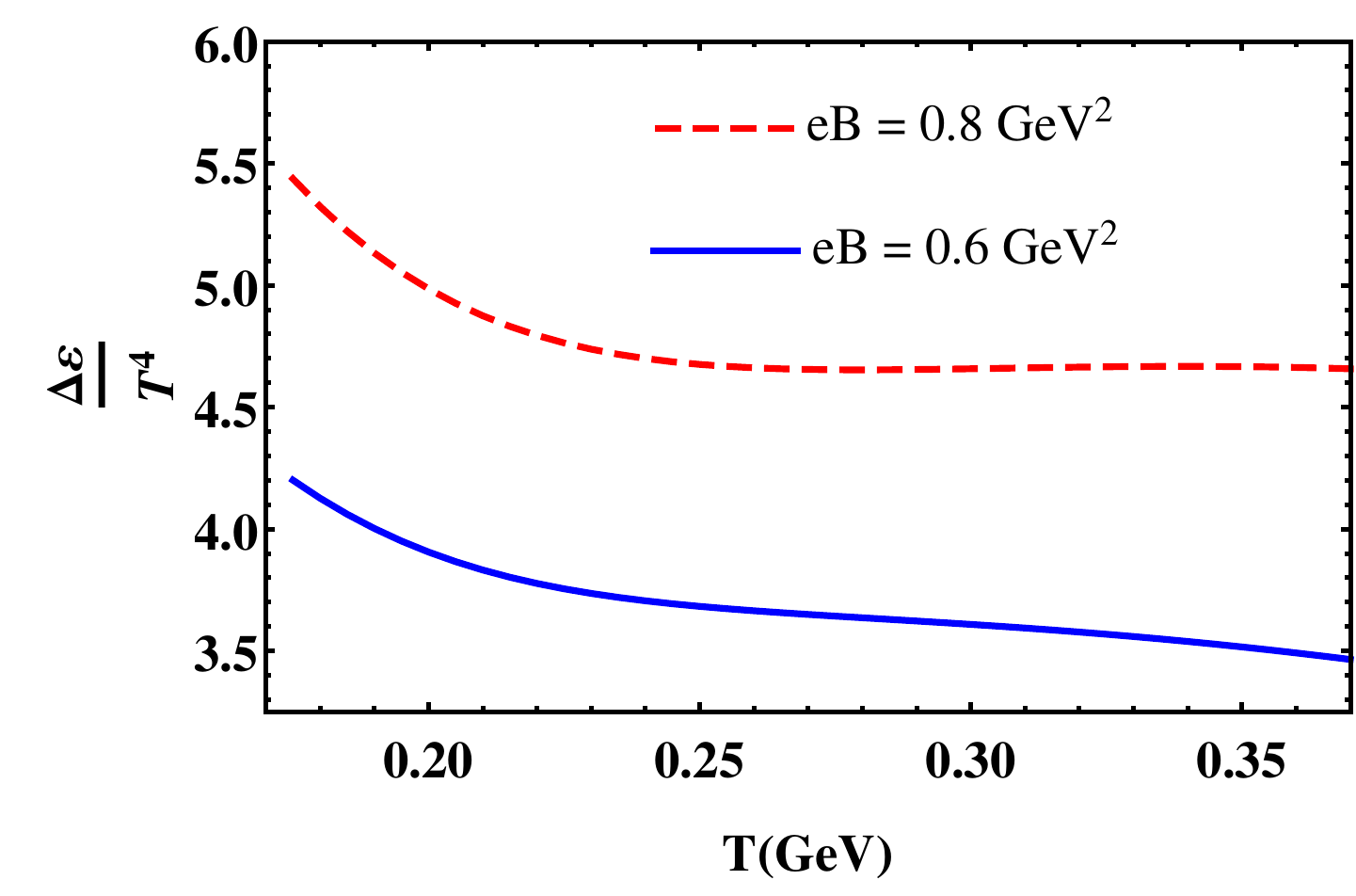}
\caption{${\Delta \varepsilon}/{T^4}$ for two different magnetic fields, plotted as a function of temperature.}
\label{fig:deltaepsilon}
\end{center}
\end{figure}
\begin{figure}
\begin{center}
\includegraphics[scale=0.6]{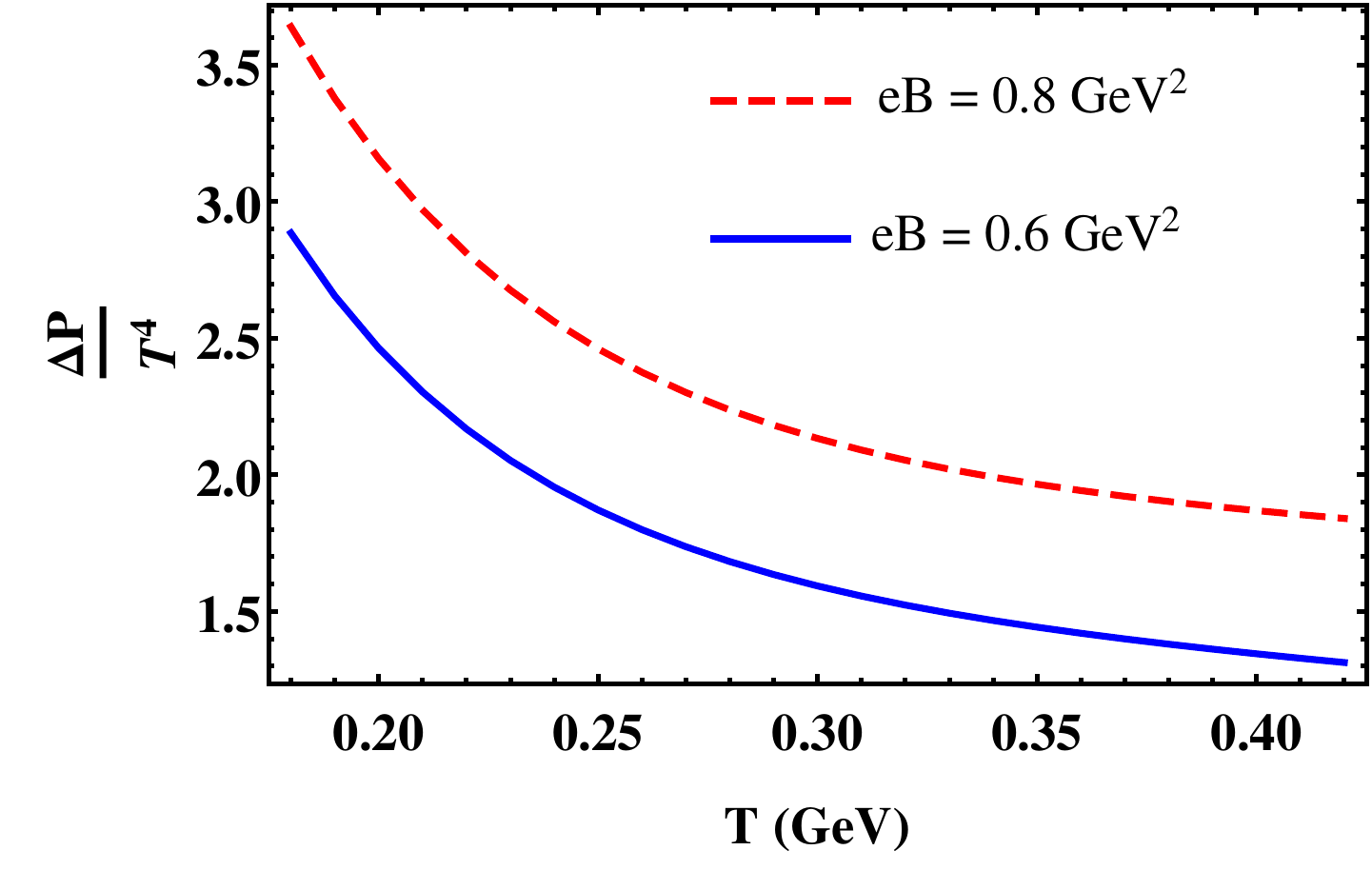}
\caption{${\Delta P}/{T^4}$ for two different magnetic fields, plotted as a function of temperature.}
\label{fig:deltapressure}
\end{center}
\end{figure}

\begin{figure}
\begin{center}
\includegraphics[scale=0.6]{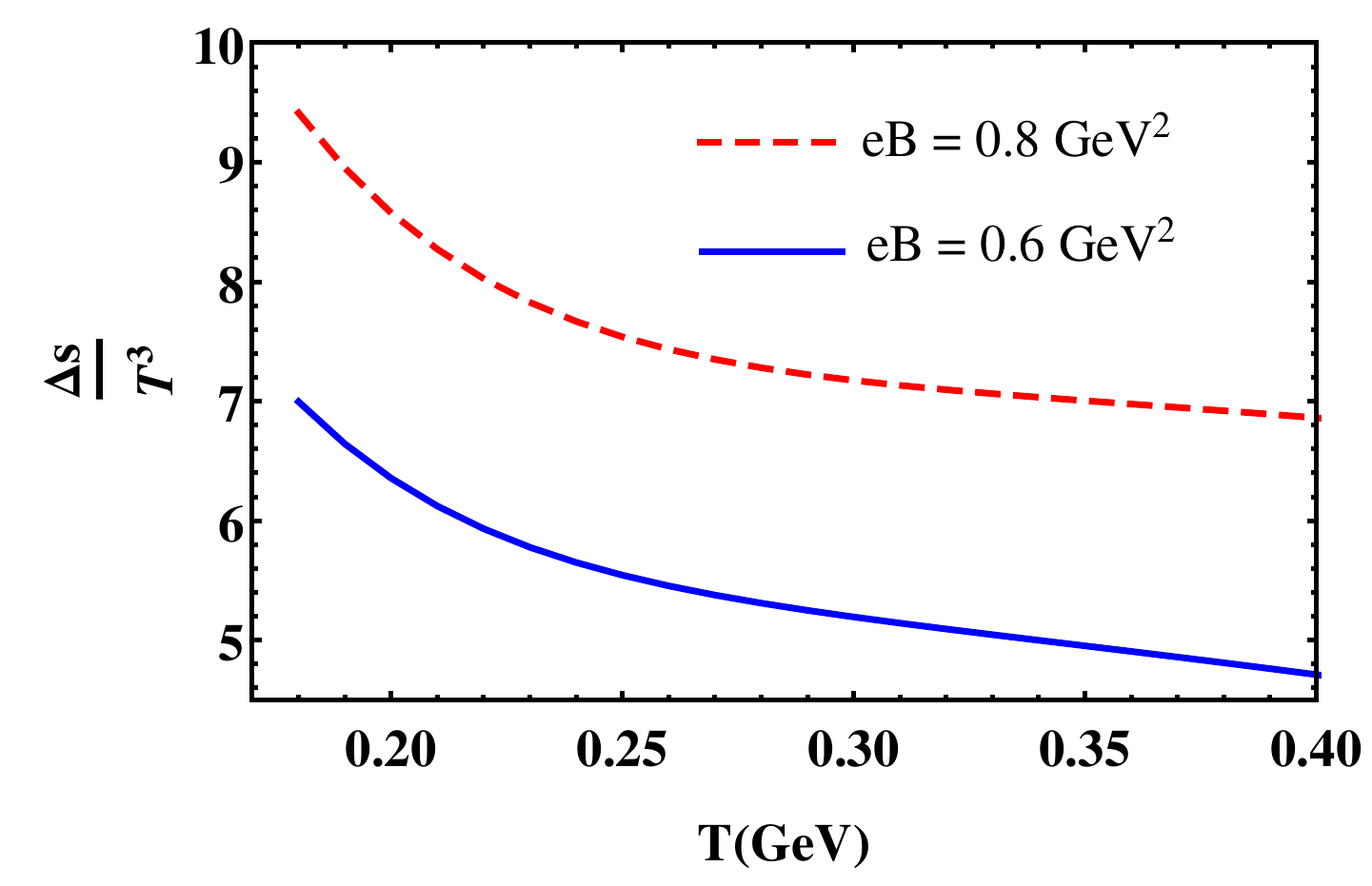}
\caption{${\Delta s}/{T^3}$ for two different magnetic fields, plotted as a function of temperature.}
\label{fig:deltaentropy}
\end{center}
\end{figure}

      We have shown the temperature dependence of different thermodynamic quantities at different values of magnetic fields in Figs. 1-3. Fig.1 shows that the energy density increases, at a given temperature, as the magnetic field increases. This is expected as, in the presence of magnetic field the total energy density goes as $\varepsilon_{total} = \varepsilon + q \textbf{M} \cdot \textbf{B}$, where $\textbf{M}$ is the magnetization \cite{Kurian2017}. We notice that the plots show correct qualitative behaviour. The quark/gluon contribution to the thermodynamic pressure, energy density, and entropy density increase with the increase in $eB$. This behaviour is consistent with that obtained using lattice QCD simulations  \cite{Bali2014} and \cite{Levkova2014}.  The same behaviour has been obtained using an effective fugacity quasiparticle model in \cite{Kurian2017}, within SU(2) NJL model in \cite{Farias2017} and using the Bag Model in \cite{Fraga2012}. There are several other studies which study magnetised quark matter. The QCD equation of state in the presence of magnetic field has been studied numerically in \cite{Bonati2013, Bonati2014}. The effect of magnetic field on QCD thermodynamics has been studied using the HTL perturbation theory both at strong\cite{Rath2017} and in weak \cite{Bandyopadhyay2017} magnetic fields. 
      
    $\Delta \varepsilon$ is the difference between energy density in the presence of magnetic fields with that in the absence of any magnetic field. This depicts the increment of energy density in the presence of the magnetic field.  The temperature dependence of $\Delta \varepsilon /T$ has been plotted in Fig.3.  In addition, we have plotted ${\Delta P}/{T^4}$ and  ${\Delta s}/{T^3}$ as functions of temperature. As expected, higher the magnetic field, higher are their values too.
    
      The transverse pressure can be obtained using equation \eqref{eq:transversepressure} and \eqref{eq:magnetization}. The calculation involves taking the derivatives with respect to the magnetic field. We are not able to calculate it here because the functional dependence of the coupling constant on the magnetic field in equation \eqref{eq:coupling} is unknown.  We emphasize that this is not a shortcoming of our model and that with the knowledge of the exact functional dependence of the coupling constant on the magnetic field we will be able to calculate the transverse pressure too.

 \section{Conclusions}We have extended the self-consistent quasiparticle model for hot QCD in the presence of magnetic fields to understand the behaviour of magnetised quark matter.  The effect of magnetic fields has been included by redefining the thermal mass of quasiparticles. The definition of thermal mass in the self-consistent model has been extended to define a thermo-magnetic mass through Landau Level quantisation for fermions. The thermodynamic quantities are evaluated by starting with the modified momentum distributions and the energy dispersion relations. The modification of these quantities has been brought about by incorporating relativistic Landau Levels.

     Using this modified quasiparticle model, we have studied the $2$-flavor system in the temperature range $170$-$400$ MeV, in the presence of magnetic fields. To this end, we made use of a parametrisation of the coupling constant that depends both on temperature and magnetic field, obtained in the context of $SU(2)$ NJL model. We found that the energy density, pressure and entropy density increase in the presence of a magnetic field as expected. Our results are qualitatively consistent with the results obtained using other approaches including lattice QCD simulations.

  The correct behaviour of the equation of state shows that the self-consistent quasiparticle model can be extended to study the thermodynamics of quark-gluon plasma in the presence of magnetic fields. For a quantitative study that can be compared with the lattice data, we need a coupling constant depending both on temperature and magnetic field.  With a proper parametrisation of the coupling constant for $2+1$ flavor, we can easily extend this work to obtain the equation of state for $2+1$ flavor QGP in the presence of magnetic field and also calculate the transverse components of pressure. We intend to do this in our future work. Another area that we plan to investigate further is how the modified equation of state affects the transport coefficients of QGP.
     \section{Acknowledgements} One of the authors Sebastian Koothottil would like to thank Manu Kurian for the helpful discussions, Arjun K. for helping with calculations and R. L. S. Farias for his comments and suggestions. Sebastian Koothottil would also like to thank UGC BSR SAP for providing research fellowship during the period of research.   
\bibliography{reference}

\end{document}